\documentclass[11pt,reqno]{amsart}
\usepackage{graphicx}

\usepackage[centertags]{amsmath}
\usepackage{amsfonts,amssymb}
\usepackage{times}
\usepackage{subfigure}
\usepackage{cite}
\usepackage{verbatim}

\newcommand{\RR}{\mathbb{R}}
\newcommand{\NN}{\mathbb{N}}

\newcommand{\bd}[1]{\boldsymbol{#1}}

\newcommand{\abs}[1]{\lvert#1\rvert}

\newcommand{\average}[1]{\langle#1\rangle}

\newcommand{\<}{\langle}
\renewcommand{\>}{\rangle}

\def\ab{\boldsymbol{a}}
\def\bb{\boldsymbol{b}}
\def\hb{\boldsymbol{h}}
\def\nb{\boldsymbol{n}}
\def\xb{\boldsymbol{x}}
\def\pb{\boldsymbol{p}}

\newcommand{\thetab}{\boldsymbol{\theta}}
\newcommand{\Thetab}{\boldsymbol{\Theta}}

\def\etab{\boldsymbol{\eta}}

\def\Wb{\boldsymbol{W}}

\newcommand{\barint}{\kern4pt \raise3.4pt\hbox{\vrule height.6pt
    width7pt} \kern-11pt \int}

\begin{document}

\title[Exact dynamical coarse-graining without time-scale separation]{
  Exact dynamical coarse-graining\\ without time-scale separation}

\author{Jianfeng Lu} 
\address{Department of
  Mathematics, Physics, and Chemistry, Duke University, Box 90320,
  Durham, NC 27708} 
\email{jianfeng@math.duke.edu}
\author{Eric Vanden-Eijnden} 
\address{Courant Institute of
  Mathematical Sciences, New York University, 251 Mercer Street, New
  York, NY 10012}
\email{eve2@cims.nyu.edu}

\maketitle

\begin{abstract}
  A family of collective variables is proposed to perform exact
  dynamical coarse-graining even in systems without time scale
  separation. More precisely, it is shown that these variables are not
  slow in general but they satisfy an overdamped Langevin equation
  that statistically preserves the sequence in which any regions in
  collective variable space are visited and permits to calculate
  exactly the mean first passage times from any such region to
  another. The role of the free energy and diffusion coefficient in
  this overdamped Langevin equation is discussed, along with the way
  they transform under any change of variable in collective variable
  space. These results apply both for systems with and without
  inertia, and they can be generalized to using several collective
  variables simultaneously. The view they offer on what makes
  collective variables and reaction coordinates optimal breaks from
  the standard notion that good collective variable must be slow
  variable, and it suggests new ways to interpret data from molecular
  dynamic simulations and experiments.
\end{abstract}

\bigskip

It is often desirable to eliminate degrees of freedom in large and
complex multi-dimensional systems, and represent their dynamics via a
reduced set of coordinates, known as coarse-grained or collective
variables. From a computational perspective such a reduction is
necessary to reach the biologically relevant length and time scales
inaccessible by all-atom molecular
simulations~\cite{ayton2007multiscale,noid2013perspective}, while in
terms of modeling it permits to explain the inner working of the
system by focusing on the most salient features of its
evolution~\cite{weinan2007heterogeneous}. To give just one example,
the folding of proteins has been described by various models of
decreasing complexity, from all-atom~\cite{best2012atomistic}, to
beads and G\={o} models, to elastic
networks~\cite{clementi2008coarse,tozzini2005coarse}, all the way down
to a one-dimensional overdamped diffusion on the free energy landscape
associated with the fraction of native contacts being
formed~\cite{socci1996diffusive,best2006diffusive,krivov2006one}. Taken
together, these models not only facilitate the simulation of proteins
but they also offer simple organization principles of protein design
and
function~\cite{bryngelson1995funnels,dobson1998protein,shea2001folding}. As
these examples show, coarse-graining typically involves a drastic
reduction in dimension and a natural question is when and why does it
work. The predominant view, shaped by the Mori-Zwanzig (MZ) projection
formalism~\cite{grabert1982projection,chorin2009stochastic}, has been
that adequate collective variables must be sufficiently slow, so that
the rest of the degrees of freedom remain at equilibrium
(adiabatically slaved) with respect to them -- in this case the
evolution of the collective variables can indeed be described by a
standard Langevin equation that can e.g. be derived from the Markovian
approximation to MZ~\cite{Zwanzig:1961,hijon2010mori} or by standard
averaging theorems~\cite{pavliotis2008multiscale}. From this
perspective coarse-graining is unfortunately quite restricted: slow
collective variables simply do not exist in most cases of interest,
and the evolution of complicated dynamical systems typically span the
totality of their wide range of time scales, from their fastest to
their slowest, without any clear separation in between. For example,
the fraction of native contacts in a protein is not a slow variable in
the standard sense of the term since it is not adiabatically separated
from the rest of the degrees of freedom in the system.

An alternative view that departs from the notion that good collective
variables must be slow variables has recently emerged in the context
of activated processes and reactive
events~\cite{EVa2004,dellago2009transition,hartmann2013characterization}. The
description of these events offer similar challenges: they are
infrequent because they require many failed attempts before occurring,
but when they finally happen they typically do so quite fast. This
means that there is no slow coordinate to describe the advancement of
the reaction in general. In spite of this, the committor function,
also known as the commitment probability or p-fold, \textit{is} a good
reaction
coordinate~\cite{du1998transition,hummer2003transition,e2005transition}
that permits to explain the mechanism of the reaction and give exact
expressions for its rate~\cite{EVa2006,Va2006,EVa2010}. Even though it
is not a slow variable, the special properties of the committor
suggest that it may be useful in the context of dynamical
coarse-graining as well. This idea was exploited in the context of
Elber's milestoning procedure~\cite{Elber:04}, in which the original
dynamics is reduced to independent transitions between hypersurfaces
(the milestones): in~\cite{VaVeCiEl2008,VaVe2008}, it was shown that
mean first passage times between these milestones can be calculated
exactly from Markovian milestoning as long as we use isocommittor
surfaces for them.  More recently Berezhkovskii and
Szabo~\cite{BerezhkovskiiSzabo:13} (see
also~\cite{krivov2012reaction}) wrote a closed, one-dimensional
Fokker-Planck (diffusion) equation whose probability flux through the
isocommittor surfaces is conserved and always equal to the exact
reaction rate.

The aim of the present communication is to elaborate on the statements
made in~\cite{BerezhkovskiiSzabo:13}, and introduce a class of
collective variables, related to the committor, to perform dynamic
coarse-graining. These variables are not slow in general, and they
depart from the standard committor in that there are not directly
connected to a reaction -- in fact, as we will see, there are no
reactant nor product states \textit{per~se} in our construction, and
infinitely many different collective variables of the type we consider
can be introduced in any given system. Yet, as we will show, for any
collective variable in this class we can write down a closed
overdamped Langevin equation that permits to calculate exactly the
mean first passage times between any two regions in which this
collective variable takes constant values. This overdamped Langevin
equation can, via specific transformations that preserve the
isosurfaces of the collective variable but relabel their values, be
written either as a driftless overdamped equation, whose associated
Fokker-Planck equation has the same form as that derived by
Berezhkovskii and Szabo~\cite{BerezhkovskiiSzabo:13}, or as a standard
overdamped equation whose coefficients involve the gradient of the
free energy and a specific diffusion coefficient. As we will see,
these results apply both to systems with and without inertia, and they
can be generalized to vector-valued (i.e. multidimensional) collective
variables. These results offer a new view on dynamical coarse-graining
that gives a criterion for optimality of reaction coordinates and
collective variables complementary to those proposed
in~\cite{krivov2006one,krivov2012reaction,berezhkovskii2004one,BerezhkovskiiSzabo:13}. They
also shed light on the dynamical meaning of the pair free
energy/diffusion coefficient that do not rely on the Markovian
approximation to MZ.

We will consider first a system whose evolution is governed by the
overdamped Langevin equation (the generalization to systems with
inertia is considered below)~\footnote{%
  Mathematically, \eqref{eq:SDE} and \eqref{eq:SDETheta} should be
  interpreted as the Ito stochastic differential equations (SDE)
\begin{displaymath}
  \begin{aligned}
    d\xb(t) & = - \beta D(\xb(t)) \nabla V(\xb(t))dt + \nabla\cdot D(\xb(t))dt\\
    & \quad + \sqrt{2}\, D^{1/2}(\xb(t))\, d\Wb(t),\\
    d\Theta(t) &= (L \theta)(\xb(t))dt + \sqrt{2} \nabla \theta(\xb(t))
    \cdot D^{1/2}(\xb(t))d \Wb(t),
  \end{aligned}
\end{displaymath}
where $\Wb(t)$ is a $N$-dimensional Wiener process. }
\begin{equation}
  \label{eq:SDE}
  \begin{aligned}
    \dot\xb(t) & = - \beta D(\xb(t)) \nabla V(\xb(t)) + \nabla\cdot D(\xb(t))\\
    & \quad + \sqrt{2}\, D^{1/2}(\xb(t))\, \etab(t),
  \end{aligned}
\end{equation}
where $\xb(t)=(x_1(t), \ldots,x_N(t))^T \in\RR^N$ denotes the
instantaneous position of the system, $V(\xb)$ the potential, $D(\xb)$
the diffusion tensor, $\beta=1/(k_BT)$ the inverse temperature, and
$\etab(t)= (\eta_1(t), \ldots,\eta_N(t))^T$ is a $N$-dimensional
white-noise process satisfying $\<\eta_i(t)\> =0$,
$\<\eta_i(t)\eta_j(s)\>= \delta_{i,j} \delta(t-s)$.  Associated
with~\eqref{eq:SDE} is the Fokker-Planck (diffusion) equation for the
probability density function $\rho(\xb,t) $ of $\xb(t)$, which
reads
\begin{equation}
  \label{eq:1}
    \frac{\partial}{\partial t} \rho(\xb,t) = \nabla \cdot \left( e^{-\beta V(\xb)} 
   D(\xb) \nabla  \bigl(e^{\beta V(\xb)}\rho(\xb,t) \bigr)\right).
\end{equation}
The stationary solution to this equation is the Boltzmann-Gibbs
(canonical) density, which is also the equilibrium probability density
function of~\eqref{eq:SDE}:
\begin{equation}
  \label{eq:BG}
  \rho_e(\xb) = Z^{-1} e^{-\beta V(\xb)},
\end{equation}
where $Z = \int_{\RR^N} e^{-\beta V(\xb)} d\xb$ is a normalization
factor. If we now introduce a (dimensionless) scalar-valued collective
variable, $\theta: \RR^N \to \RR$, a simple application of Ito's
lemma~\cite{bass1998diffusions} indicates that $\Theta(t) \equiv
\theta(\xb(t))$ satisfies
\begin{equation}
  \label{eq:SDETheta}
  \dot\Theta(t) = (L \theta)(\xb(t)) 
  + \sqrt{2} \nabla \theta(\xb(t)) \cdot D^{1/2}(\xb(t))\etab(t), 
\end{equation}
where the generator $L$ is the adjoint of the operator at the right
hand side of~\eqref{eq:1} and its action on $\theta(\xb)$ reads
\begin{equation}
  \label{eq:2}
  (L\theta)(\xb) \equiv e^{\beta V(\xb)} \nabla \cdot \bigl(
    e^{-\beta V(\xb)} D(\xb) \nabla \theta(\xb) \bigr).
\end{equation}
Since the right hand side of~\eqref{eq:SDETheta} depends on $\xb(t)$
rather than $\Theta(t)$ alone, this equation is not closed -- this is
the issue of dynamical coarse-graining made explicit~\footnote{%
  Note that~\eqref{eq:SDETheta} can be closed for any $\theta(\xb)$
  using time-dependent averaging conditional on $\Theta(t) =
  \theta(\xb(t))$ using $\rho(\xb,t)/\int_{\RR^N}\rho(\xb,t) \delta
  (\theta(\xb) - \Theta(t))d\xb$ (rather than the standard equilibrium
  conditional averaging using $\rho_e(\xb)/\int_{\RR^N}\rho_e(\xb)
  \delta (\theta(\xb) - \Theta(t))d\xb$). The result, however, is a
  non-equilibrium evolution equation with time-dependent
  coefficients. For details see~\cite{LeLe2010}.}.

\begin{figure}
   \centerline{\includegraphics[scale=0.3]{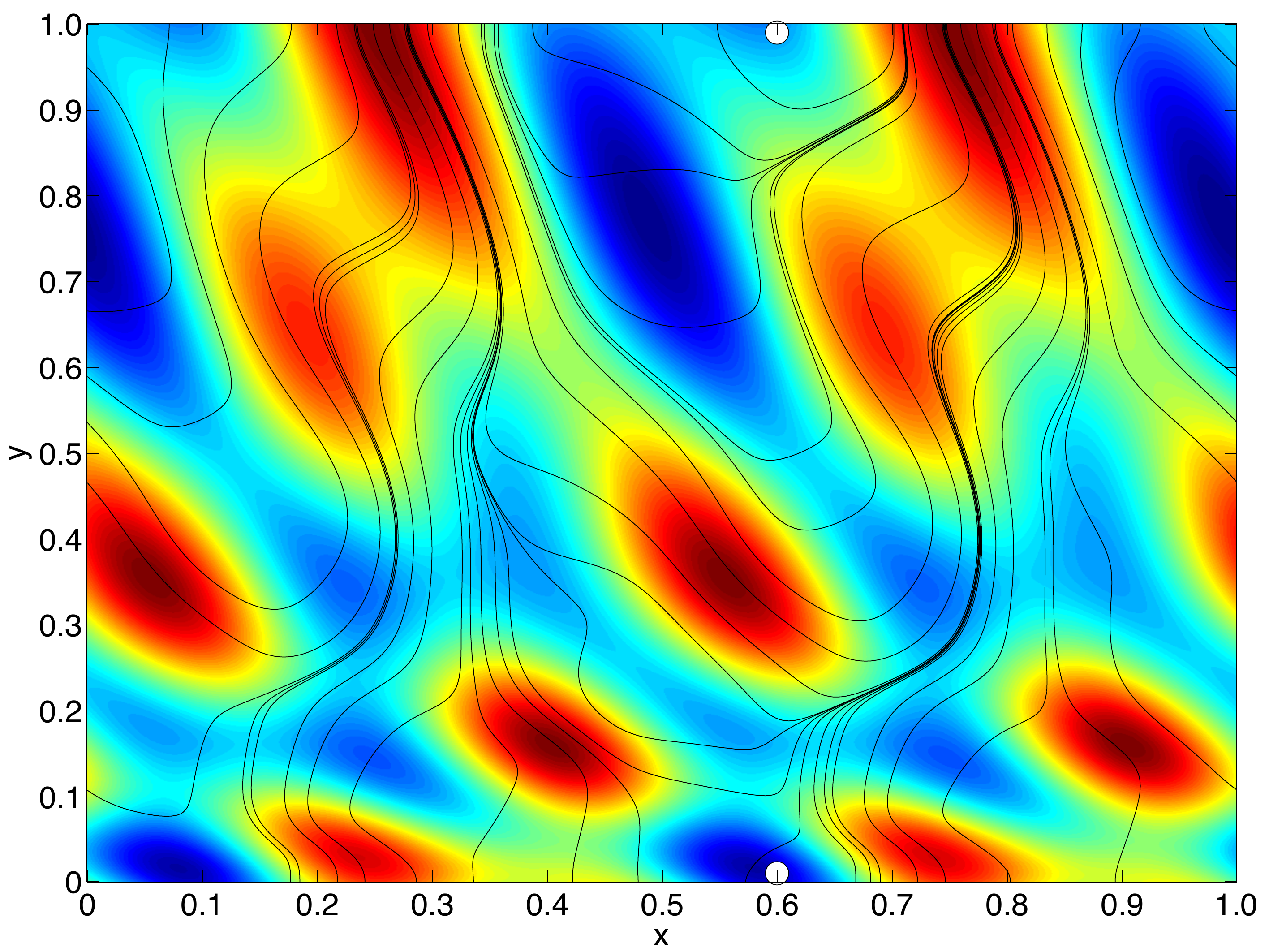}}
   \caption{Illustration of the potential solution of~\eqref{eq:sbke}
     for a two-dimensional toy potential shown in the background --
     this potential is periodic in the $x$-direction, but not in the
     $y$-direction. We solve~\eqref{eq:sbke} with $\beta=1$,
     $D(\xb)=\text{Id}$ and by imposing periodic boundary conditions
     in $x$ and no flux (Neumann) boundary conditions at $y=0$ and
     $y=1$. The two points $\ab=(0.01,0.6)$ and $\bb=(0.99,0.6) $ are
     shown as white dots. The isolines of the potential
     $\theta(\xb)=\theta(x,y)$ are shown in black: note how they
     follow the features of the potential. }
   \label{fig:1}
\end{figure}

To proceed further, let us introduce a specific class of collective
variables $\theta(\xb)$ via the solution to
\begin{equation}
  \label{eq:sbke}
  \nabla \cdot (\rho_e(\xb) D(\xb) \nabla \theta(\xb)) = 
  \tau \bigl(\delta(\xb - \ab) - \delta(\xb - \bb)\bigr), 
\end{equation}
where $\tau$ is an arbitrary time scale introduced for dimensional
consistency, and $\ab\in\RR^N$ and $\bb\in\RR^N$ are two arbitrary
points in configuration space. By varying the location of these
points, different $\thetab(\xb)$ can be defined that each can be
thought of as the potential associated with the pair of opposite point
charges at $\ab$ and $\bb$, with $\rho_e(\xb) D(\xb)$ playing the role
of dielectric~\footnote{%
  Under additional assumptions, we could introduce the potential
  associated with a single point charge, rather than the pair at $\ab$
  and $\bb$, or with different distribution of charges. We chose to
  work with \eqref{eq:sbke} for simplicity: notice in particular that
  the Fredholm's alternative guarantees that this equation has a
  unique solution since $\int \bigl( \delta(\xb-\ab) -
  \delta(\xb-\bb)\bigr) d\xb = 0$.}. Note that the solution
to~\eqref{eq:sbke} maps $\RR^N\setminus\{\ab,\bb\}$ onto
$(-\infty,\infty)$ and has no extrema in its domain, which makes
$\theta(\xb)$ suitable as a collective variable -- see
Fig.~\ref{fig:1} for an illustration on a two-dimensional
example. Note also that \eqref{eq:sbke} can be rewritten using the
generator $L$ as
\begin{equation}
  \label{eq:sbke2}
  (L \theta)(\xb) = \tau \rho^{-1}_e(\xb) \bigl(\delta(\xb -\ab) 
  - \delta(\xb - \bb) \bigr), 
\end{equation}
which implies that $L \theta = 0$ at every point except~$\ab$
and~$\bb$. This property will be key in the developments below.  The
potential $\theta(\xb)$ is related to the committor.  To see why, let
$A$ and $B$ be the two one-parameter family of sets defined as
\begin{equation}
  \label{eq:5}
    A = \bigl\{ \xb \mid \theta(\xb) \leq \theta_A \bigr\}, \quad
    B = \bigl\{ \xb \mid \theta(\xb) \geq
    \theta_B \bigr\},
\end{equation}
where $\theta_A < \theta_B$ can take arbitrary values, and define the
function $q_{AB}(\xb)$ as~\footnote{%
  Note that $q_{AB}(\xb)$ is independent of~$\tau$, the arbitrary time
  scale introduced in the definition of~$\theta(\xb)$.}
\begin{equation}
  \label{eq:committor}
  q_{AB}(\xb) = \frac{\theta(\xb) - \theta_A}{\theta_B - \theta_A} \qquad
  \text{if \ $\xb \not \in A\cup B$} 
\end{equation}
and $q_{AB}(\xb)= 0$ if $\xb \in A$ and $q(\xb)=1$ if $ \xb \in B$.  It is
easy to verify from~\eqref{eq:sbke} that $q_{AB}(\xb)$ is the solution
to 
\begin{equation} 
  \label{standardbke} 
  (L q_{AB})(\xb) = 0,  \qquad
  \text{if \ $\xb \not \in A\cup B$},
\end{equation}
with boundary condition $q_{AB}(\xb)= 0$ if $\xb \in \partial A$ and
$q_{AB}(\xb)=1$ if $ \xb \in \partial B$. Hence, $q_{AB}(\xb)$ is just
the committor function for the transition between the `reactant'
set~$A$ and the `product' set~$B$, that is, $q_{AB}(\xb)$ gives the
probability that a trajectory starting at point $\xb$ will reach $B$
rather than $A$ next~\cite{EVa2006}. Note however that the family of
reactant and product states defined above do not need to be associated
with an actual reactive process -- in particular, they do not need to
be metastable. Also, $\theta(\xb)$ is an actual collective variable
that can be used everywhere, unlike $q_{AB}(\xb)$ that is constant
inside $A$ and $B$.

Next, let us consider the evolution of $\Theta(t) = \theta(\xb(t))$ when
$\theta(\xb)$ solves~\eqref{eq:sbke}. Since $L\theta = 0$ as long as the
dynamics stays away from the points $\ab$ and $\bb$ (which happens
with probability 1 if the space dimension is $N>1$),
\eqref{eq:SDETheta} reduces to the driftless equation
\begin{equation}
  \label{eq:eqnQ}
  \dot\Theta(t)  = \sqrt{2} \nabla
  \theta(\xb(t))\cdot D^{1/2}(\xb(t)) \etab(t).
\end{equation}
The absence of drift term in this equation has an important
consequence.  If we introduce the (random and dimensionless) rescaled time
\begin{equation}
  \label{eq:taudef}
  s(t) = \int_0^t \bigl\lvert \nabla \theta(\xb(t'))\cdot 
  D(\xb(t')) \nabla \theta(\xb(t')) \bigr\rvert dt', 
\end{equation}
then the right hand side of~\eqref{eq:eqnQ} has the same statistical
properties (i.e.~the same law or distribution) as $\dot s(t) \eta(s(t))$,
where $\eta(s)$ is a one-dimensional white-noise~\footnote{%
  Mathematically \eqref{eq:eqnQ} should be interpreted as the Ito SDE
\begin{displaymath}
  d \Theta(t)  = \sqrt{2} \nabla
  \theta(\xb(t))\cdot D^{1/2}(\xb(t)) d\Wb(t),
\end{displaymath}
and we are using the following identity in law (meaning that both
sides of the equality have the same distribution)
\begin{displaymath}
  \textstyle\int_0^t\nabla\theta(\xb(t'))\cdot D^{1/2}(\xb(t')) d\Wb(t')
  \, 
  {\buildrel d \over =}\, W(s(t)),
\end{displaymath}
where $W(s)$ is a one-dimensional Wiener process. As a result the
SDE for $\theta(t)$ can be written in terms of $s(t)$ as $d\Theta(s)
= \sqrt{2} dW(s)$, which is \eqref{eq:eqnQ2}.}. In other words, in
terms of~$s$, \eqref{eq:eqnQ} simply reads
\begin{equation}
  \label{eq:eqnQ2}
  \frac{d\Theta}{ds} = \sqrt{2} \eta(s),
\end{equation}
which, unlike~\eqref{eq:eqnQ}, \textit{is} closed. Of~course, the
simplicity of~\eqref{eq:eqnQ2} is deceptive, since the rescaled time
defined in~\eqref{eq:taudef} depends on $\xb(t)$ and is not known
explicitly. In other words, in~\eqref{eq:eqnQ2} we have lost the
physical time information about the process. Still, \eqref{eq:eqnQ2}
is a useful starting point for further developments. Indeed, the fact
that we can put~\eqref{eq:eqnQ} in the form of~\eqref{eq:eqnQ2} by
rescaling time (something that cannot be done with~\eqref{eq:SDETheta}
with a general $\theta(\xb)$ due to presence of the drift term
$(L\theta)(\xb(t))$ in this equation), means that \eqref{eq:eqnQ2}
captures exactly, if not the times, at least the order of the sequence
in which the trajectory~$\xb(t)$ visits different regions defined via
$\theta(\xb)$ -- for example, any sets of hypersurfaces defined as
$\theta(\xb)=\theta_j$, $j=1, \ldots, M$ for any $M\in \NN$ and any
values of the constants $\theta_j$. This property was already used in
the context of milestoning to introduce a set of optimal milestones
between which the sequence of transitions is
Markov~\cite{VaVeCiEl2008}. It also suggests that we might be able to
recover some kinetic information about the process by reintroducing
the time, at least in some average sense, as was also done in optimal
milestoning~\cite{VaVeCiEl2008}.  A natural procedure to do this is to
canonically average~\eqref{eq:eqnQ} conditionally on $\theta(\xb(t))=
\Theta(t)$~\footnote{%
  It is natural to average~\eqref{eq:eqnQ} to get mean first passage
  times because such first passage times are the sum of passage times
  between isosurfaces of $\theta(\xb)$, and the average of a sum is
  the sum of the averages. Note however that this also explains why
  higher order moments of the first passage times cannot, in general,
  be calculated exactly from~\eqref{eq:3}}. This gives
\begin{equation}
  \label{eq:3}
  \dot\Theta(t)  = \sqrt{2} \sigma(\Theta(t)) \eta(t), 
\end{equation}
where we defined $\sigma$ as the square root of the conditional
expectation
\begin{equation}
  \label{eq:defM}
  \begin{aligned}
    &\sigma^2(\theta) = \bigl\langle \, \nabla \theta(\xb)\cdot D(\xb)
    \nabla
    \theta(\xb) \mid \theta(\xb) = \theta \, \bigr\rangle \\
    &\ = \frac{\int_{\RR^N} \nabla \theta(\xb)\cdot D(\xb) \nabla
      \theta(\xb) \rho_e(\xb) \delta(\theta(\xb) - \theta) d\xb}
    {\int_{\RR^N} \rho_e(\xb) \delta(\theta(\xb) - \theta) d\xb}.
  \end{aligned}
\end{equation}
\eqref{eq:3} is not equivalent to~\eqref{eq:eqnQ} (i.e. it is not
exact), but it permits to calculate \textit{exactly} the mean first
passage times taken by trajectory of the original process $\xb(t)$ to
travel between any two regions defined via $\theta(\xb)$ -- for
example, between any two hypersurfaces defined as
$\theta(\xb)=\theta_A$ and $\theta(\xb)=\theta_B$. A similar statement
was also made by Berezhkovskii and Szabo~\cite{BerezhkovskiiSzabo:13}
-- here we will prove it by using the connection between the potential
$\theta(\xb)$ and the committor function, and using results from
transition path theory (TPT)~\cite{EVa2006,Va2006,EVa2010}. Before
doing so, however, let us rewrite~\eqref{eq:3} is a way that makes
apparent the connection with the results
in~\cite{BerezhkovskiiSzabo:13}.

To this end note that $\sigma^2(\theta)$ can also be written as
\begin{equation}
  \label{eq:defsigma}
  \sigma^2(\theta) = \nu  e^{\beta G(\theta)},
\end{equation}
where $G(\theta)$ is the free energy associated with $\theta(\xb)$, 
\begin{equation}
  \begin{aligned}
    G(\theta) & = - \beta^{-1} \ln \< \delta(\theta(\xb) -
    \theta)\> \\
    & \equiv - \beta^{-1} \ln \int_{\RR^N} \rho_e(\xb)
    \delta(\theta(\xb) - \theta) d\xb,
  \end{aligned}
\end{equation}
and we defined
\begin{equation}
  \label{eq:4}
  \nu = \int_{\RR^N}  \nabla \theta(\xb)\cdot D(\xb) \nabla
  \theta(\xb) \rho_e(\xb) \delta(\theta(\xb) - \theta) d\xb.
\end{equation}
This factor is a constant (independent of $\theta$) as can be seen by
taking its derivative of $\nu$ with respect to $\theta$:~\footnote{%
  Mathematically, the way to prove that $\nu$ is constant is to use
  the co-area formula to express it as the surface integral
\begin{displaymath}
  \textstyle \nu  = \int_{S_\theta} \hat{\nb}_{S_\theta}(\xb) \cdot D(\xb) \nabla \theta(\xb)
  \rho_e(\xb) d\sigma(\xb)
\end{displaymath}
where $S_\theta = \{\, \xb \mid \theta(\xb) = \theta\, \}$, $\hat{\nb}_{S_\theta}$
is the unit normal pointing in direction of increasing $\theta(\xb)$ and
$d\sigma(\xb)$ is the Hausdorff measure on $S_\theta$.  By the divergence
formula, for any $\theta_A < \theta_B$, we then have
\begin{displaymath}
  \begin{aligned}
    &\textstyle \int_{S_{\theta_A}} \hat{\nb}_{S_{\theta_A}}(\xb) \cdot
    D(\xb) \nabla \theta(\xb)
    \rho_e(\xb) d\sigma(\xb) \\
    & - \textstyle \int_{S_{\theta_B}}\hat{\nb}_{S_{\theta_B}}(\xb) \cdot
    D(\xb) \nabla \theta(\xb)
    \rho_e(\xb) d\sigma(\xb) \\
    & = \textstyle \int_{\Omega_{AB}} \nabla \cdot ( \rho_e(\xb)
    D(\xb) \nabla \theta(\xb)) d \xb = 0.
  \end{aligned}
\end{displaymath}
where $\Omega_{AB} = \{\, x \mid \theta_A < \theta(\xb) < \theta_B\,\}$.
}
\begin{displaymath}
  \begin{aligned}
    & \int_{\RR^N} \nabla \theta(\xb)\cdot D(\xb) \nabla
    \theta(\xb) \rho_e(\xb) \delta'(\theta(\xb) - \theta) d\xb\\
    & = - \int_{\RR^N} \rho_e(\xb) \nabla \theta(\xb)\cdot D(\xb)
    \nabla \delta(\theta(\xb) - \theta)d\xb\\
    & = \int_{\RR^N} \nabla\cdot \left(\rho_e(\xb) D(\xb) \nabla
      \theta(\xb) \right) \delta(\theta(\xb) - \theta)d\xb= 0,
  \end{aligned}
\end{displaymath}
where we used the chain rule to get the first equality, integration by
parts to get the second, and \eqref{eq:sbke} to get the
third. Using~\eqref{eq:defsigma} in~\eqref{eq:3}, we see that the
Fokker-Planck equation for the probability density of $\Theta(t)$
solution to~\eqref{eq:3} is
\begin{equation}
  \label{eq:6}
  \frac{\partial}{\partial t} \bar\rho(\theta,t) = \nu
  \frac{\partial^2}{\partial\theta^2} \left( e^{\beta G(\theta)} \bar\rho(\theta,t)\right),
\end{equation}
which is essentially a rewriting of Eq.~(3.6)
in~\cite{BerezhkovskiiSzabo:13}, the only difference being that we
wrote~\eqref{eq:6} (and~\eqref{eq:3}) using the potential
$\theta(\xb)$ rather than the committor function -- this is because we
want~\eqref{eq:3} (and~\eqref{eq:6}) to be defined everywhere,
which is the case if we use $\theta(\xb)$, but not the committor
(since this function is constant inside the reactant and product
states). In particular, Eq.~(3.6) in~\cite{BerezhkovskiiSzabo:13}
needs boundary conditions at the reactant and product states, whereas
\eqref{eq:6} does not. Note also that the stationary solution
to~\eqref{eq:6} is $\bar\rho_e(\theta) = e^{-\beta G(\theta)}$ as it
should be.

Let us now justify the claim that~\eqref{eq:3} permits to calculate
mean first passage times exactly. We recall from TPT that the
statistical properties of the reactive trajectories (that is, the
pieces of trajectories during which they transition from $A$ to $B$
without any return to $A$ along the way) can be expressed in terms of
$\rho_e(\xb)$ and $q_{AB}(\xb)$.  In particular, the reaction rate
from $A$ to $B$ (that is, the average number of reactive trajectories
observed per unit of time) can be calculated as
\begin{equation}
  \label{eq:reactionrate} 
  \begin{aligned}
    \nu_{AB} &= \int_{(A \cup B)^c} \nabla q_{AB}(\xb)\cdot D(\xb)
    \nabla q_{AB}(\xb)
    \rho_e(\xb) d\xb \\
    &= \int_{(A \cup B)^c} \frac{\nabla \theta(\xb)\cdot D(\xb) \nabla
      \theta(\xb)}{(\theta_B-\theta_A)^2}
    \rho_e(\xb) d\xb \\
    &= \frac1 {(\theta_B-\theta_A)^2} \int_{\theta_A}^{\theta_B}
    \int_{\RR^N} \nabla \theta(\xb)\cdot
    D(\xb) \nabla \theta(\xb)\\
    &\qquad\qquad \qquad\qquad \times
    \rho_e(\xb) \delta(\theta(\xb) - \theta)d\xb d\theta \\
    &= \frac{\nu}{\theta_B - \theta_A},
  \end{aligned}
\end{equation}
where we started from the result from TPT, then
used~\eqref{eq:committor}, \eqref{eq:4}, and the constancy of~$\nu$.
Similarly the mean first passage time from $A$ to $B$ (that is, the
average time it takes to return to $B$ after hitting $A$ the first
time after leaving $B$) is given by
\begin{equation}
  \label{eq:tauab}
  \begin{aligned}
    \tau_{AB} & = \nu_{AB}^{-1} \int_{\RR^N} \rho_e(\xb) ( 1 -
    q_{AB}(\xb)) d\xb\\
    & = \nu^{-1}\int_{(A \cup B)^c} \rho_e(\xb) ( \theta_B -
    \theta(\xb)) d\xb\\
    & \quad + \nu^{-1} ( \theta_B -
    \theta_A)\int_{A} \rho_e(\xb) d\xb\\
    & = \nu^{-1} \int_{\theta_A} ^{\theta_B} (\theta_B - \theta) e^{-\beta
      G(\theta) } d\theta \\
    & \quad + \nu^{-1} ( \theta_B - \theta_A)\int_{-\infty}^{\theta_A}
    e^{-\beta G(\theta) } d\theta.
  \end{aligned}
\end{equation}
This last formula justifies our claim: indeed the mean first passage
time $\tau_B(\theta)$ from any $\theta<\theta_B$ to $\theta_B$ of the
solution to~\eqref{eq:eqnQ2} solves~\cite{bass1998diffusions}
\begin{equation}
  \label{eq:7}
  \begin{aligned}
    \nu e^{\beta G(\theta)} \frac{d^2\tau_B}{d\theta^2} = -1, \
    \tau_B(\theta_B) =0, \
    \lim_{\theta\to-\infty} \frac{d\tau_B}{d\theta} =0.
  \end{aligned}
\end{equation}
It is easy to see that the solution to this equation evaluated at
$\theta=\theta_A$ coincide with~\eqref{eq:tauab}, $\tau_B(\theta_A)
\equiv \tau_{AB}$. Since $\theta_A$ and $\theta_B>\theta_A$ are
arbitray in this argument, and we can easily generalize it to the case
with $\theta_B<\theta_A$, we can indeed calculate exactly mean first
passage times of the original process $\xb(t)$ from any surface
$\theta(\xb) = \theta_A$ to any any surface $\theta(\xb) = \theta_B$
using~\eqref{eq:3}.

One thing still remain to be done, namely show that~\eqref{eq:3} can
be recast into a (or rather infinitely many, all equivalent) standard
overdamped Langevin equation(s). Clearly, we can change the form
of~\eqref{eq:3} without affecting the physics behind this equation by
any change of variable, i.e. by introducing $\tilde\Theta(t) =
h(\Theta(t))$ for any monotonic (one-to-one) function $h$ mapping
$\RR$ onto $\RR$. This corresponds to using $\tilde\theta(\xb) =
h(\theta(\xb))$ as new collective variable, and it turns~\eqref{eq:3}
into~\footnote{%
  Notice that \eqref{eq:11} holds for any choice of $h(\theta)$:
  however, for the specific choice $ h'(\theta) = \exp(-\tfrac12\beta
  G(\theta))$, we have $m(\tilde \theta) = \nu$, and so \eqref{eq:11}
  reduces to
\begin{displaymath}
  \Dot{\tilde\Theta}(t) = - \beta \nu \tilde G'(\tilde\Theta(t)) + \sqrt{2 \nu}\, \eta(t),
\end{displaymath}
which is an overdamped equation with a constant diffusion
coefficient~$\nu$. While simpler than~\eqref{eq:11} with a
$\tilde\theta$-dependent $m(\tilde\theta)$, this equation is not
better -- it is completely equivalent to~\eqref{eq:11}. Note also
that, in the vector-valued case, no change of variable permits to
turn~\eqref{eq:vectval} into an equation in which the diffusion tensor
is constant.}
\begin{equation}
  \label{eq:11}
  \begin{aligned}
    \Dot{\tilde \Theta}(t) & = -\beta m(\tilde \Theta (t)) \tilde
    G'(\tilde\Theta (t)) + m'(\tilde \Theta (t))\\
    & \qquad\qquad\qquad\qquad + \sqrt{2} m^{1/2}(\tilde \Theta (t))\, \eta(t),
  \end{aligned}
\end{equation}
where $\tilde G(\tilde \theta) $ is the free energy associated with $\tilde
\theta(\xb)$, 
\begin{equation}
  \label{eq:FE}
  \tilde G(\tilde\theta) = - \beta^{-1} \ln \<\delta(\tilde\theta(\xb) - \tilde\theta)\>,
\end{equation}
and the diffusion coefficient~$m(\tilde\theta)$ is given
by~\footnote{%
  Note that $m(\tilde \theta)$, like $\nu$, has the dimension of the
  inverse of a time since $\theta(\xb)$ is dimensionless.}
\begin{equation}
  \label{eq:12}
  m(\tilde\theta) = \bigl\langle \nabla \tilde\theta(\xb)\cdot D(\xb) \nabla \tilde\theta(\xb) \mid
  \tilde\theta(\xb) = \tilde\theta  \bigr\rangle.
\end{equation}
In fact, it is easy to see that~\eqref{eq:3} itself is in the form
of~\eqref{eq:11}, with $m(\theta) \equiv \sigma^2(\theta) = \nu
e^{\beta G(\theta)}$, which implies that $-\beta m(\theta) G'(\theta)
+ m'(\theta) = 0$.  Since \eqref{eq:11} contains the same physics
as~\eqref{eq:3} it can again be used to calculate exactly mean first
passage times from any surface where $\tilde\theta(\xb) =
\tilde\theta_A$ to any from any surface where $\tilde\theta(\xb) =
\tilde\theta_B$.  To derive~\eqref{eq:11}, notice first that Ito's
lemma implies that $\tilde\Theta(t)= h(\Theta(t))$ satisfies
\begin{equation}
  \label{eq:coarse}
  \Dot{\tilde\Theta}(t) = \sigma^2(\Theta(t)) h''(\Theta(t))  
  + \sqrt{2} h'(\Theta(t)) \sigma(\Theta(t)) \eta(t).
\end{equation}
To cast this equation in the form~\eqref{eq:11}, by looking at the
noise term we see that we must take
\begin{equation}
  \label{eq:rel1}
  m(h(\theta)) = |h'(\theta)|^2 \sigma^2(\theta),
\end{equation} 
On the other hand the free energies $G(\tilde\theta) $ and $G(\theta)$
are related as
\begin{displaymath}
  \begin{aligned}
    \tilde G(h(\theta)) & = - \beta^{-1} \ln \int \rho_e(\xb) \delta(\tilde\theta(\xb) - h(\theta))
    d\xb \\
    & = - \beta^{-1} \ln \int \rho_e(\xb) \delta(h(\theta(\xb)) -
    h(\theta)) d\xb \\
    & = - \beta^{-1} \ln \int \rho_e(\xb) \abs{h'(\theta)}^{-1} 
    \delta(\theta(\xb) - \theta) d\xb \\
    & = - \beta^{-1} \ln \bigl(\abs{h'(\theta)}^{-1} e^{-\beta G(\theta)}
    \bigr) \\
    & =  G(\theta) + \beta \ln |h'(\theta)|. 
  \end{aligned}
\end{displaymath}
Together with~\eqref{eq:defsigma}, this relationship implies that
\begin{displaymath}
  \begin{aligned}
    \sigma^2(\theta) \equiv \nu e^{\beta G(\theta)} =\frac{\nu e^{\beta
        \tilde G(h(\theta))}}{h'(\theta)},
  \end{aligned}
\end{displaymath}
which we can combine with~\eqref{eq:rel1} to get
\begin{equation}
  \label{eq:rel2}
  m(h(\theta)) =  \nu  e^{\beta
    \tilde G(h(\theta))}h'(\theta).
\end{equation}
Solving this equation in $h'(\theta)$, then differentiating over
$\theta$ and multiplying by $\sigma^2(\theta) = \nu e^{-\beta
  \tilde G(h(\theta))}/h'(\theta)$ gives
\begin{equation}
  \label{eq:10}
  \sigma^2(\theta) h''(\theta) = m'(h(\theta))- \beta m(h(\theta)) G'(h(\theta)).
\end{equation}
This shows that the drift term in~\eqref{eq:coarse} is also equal to
that in~\eqref{eq:11}. Finally, to show that $m(\tilde \theta)$ is
given by~\eqref{eq:12}, use the definition~\eqref{eq:4} of $\nu$
in~\eqref{eq:rel2} to get
\begin{displaymath}
  \begin{aligned}
    m(h(\theta)) & = e^{\beta \tilde G(h(\theta))}
    h'(\theta)\int_{\RR^N} \nabla \theta(\xb)\cdot D(\xb) \nabla
    \theta(\xb)\\
    &\qquad\quad \times \rho_e(\xb) \delta(\theta(\xb) - \theta) d\xb\\
    & = e^{\beta \tilde G(h(\theta))}
    \int_{\RR^N} \nabla h(\theta(\xb))\cdot D(\xb) \nabla
    h(\theta(\xb))\\
    &\qquad\quad \times \rho_e(\xb) \delta(h(\theta(\xb)) - h(\theta)) d\xb.
  \end{aligned}
\end{displaymath}
Note that even if $\Theta(t)$ solves~\eqref{eq:3}, in order for
\eqref{eq:11} to preserves the right physics for any $\tilde \Theta(t)
= h(\Theta(t))$ both the free energy $G(\tilde\theta)$ and the
diffusion coefficient~$m(\tilde \theta)$ must be changed consistently:
in other words, the pair $\tilde G(\tilde\theta)$, $m(\tilde \theta)$
rather than the free energy alone carries dynamical
meaning~\cite{e2005transition,maragliano2006string,best2010coordinate,berezhkovskii2011time}. Our
results generalizes this observation to situations without time-scale
separation, as long as a potential $\theta(\xb)$ solution
of~\eqref{eq:sbke} is used as collective variables.

Our results can be generalized to systems with inertia,
e.g. when~\eqref{eq:SDE} is replaced by the Langevin equation
\begin{equation}
  \label{eq:SDEL}
  m \ddot \xb + \gamma(\xb) \dot\xb = - \nabla V(\xb) 
  + \sqrt{2\beta^{-1}}\, \gamma^{1/2}(\xb)\, \etab(t),
\end{equation}
where $m$ is the mass matrix and $\gamma(\xb)$ is the friction tensor,
related to the diffusion tensor $D(\xb)$ in~\eqref{eq:SDE} via
Einstein's relation: $\gamma(\xb)= \beta^{-1} D^{-1}(\xb)$. Then, the
potential $\theta(\xb)$ becomes a function $\vartheta(\xb,\pb)$ of
both positions~$\xb$ and momenta~$\pb = m^{-1} \dot \xb$ and satisfies
(compare~\eqref{eq:sbke})
\begin{equation}
  \label{eq:sbkeL}
  \begin{aligned}
    &\beta^{-1}(\partial_{\xb}\ \partial_{\pb}) \cdot \left( \varrho_e(\xb,
      \pb) \begin{pmatrix} 0 & \text{Id} \\ -\text{Id} &
        \gamma(\xb) \end{pmatrix}
      \begin{pmatrix} \partial_{\xb} \\ \partial_{\pb} \end{pmatrix}
      \vartheta(\xb, \pb)\right) \\
    &= \tau \bigl(\delta(\xb-\xb_a)\delta(\pb- \pb_a)
    -\delta(\xb-\xb_b)\delta(\pb- \pb_b) \bigr).
  \end{aligned}
\end{equation}
Here $(\xb_a,\pb_a)$ and $(\xb_b,\pb_b)$ are two arbitrary points in
phase-space, and $\varrho_e(\xb, \pb)$ is the equilibrium (canonical)
probability density of~\eqref{eq:SDEL}:
\begin{equation}
  \label{eq:8}
  \varrho_e(\xb, \pb) = \mathcal{Z}^{-1} e^{-\beta H(\xb,\pb)}
\end{equation}
where $H (\xb,\pb)= \tfrac12 \pb \cdot m^{-1} \pb + V(\xb)$ and
$\mathcal{Z}$ is the partition function. If we use $\vartheta(\xb,\pb)$ as
collective variable, the results obtained in the overdamped case can
be straightforwardly generalized to the present situation. In
particular if we set $\Theta(t) = \vartheta(\xb(t),\pb(t))$ and use the
rescaled time (compare~\eqref{eq:taudef})
\begin{equation}
  \label{eq:9}
  \begin{aligned}
    s(t) = \beta^{-1} \int_0^t & |\partial_{\pb} \vartheta(\xb(t'),\pb(t'))\\
    & \cdot \gamma(\xb(t')) \partial_{\pb} \vartheta(\xb(t'),\pb(t'))| dt',
  \end{aligned}
\end{equation}
then $\Theta(s)$ satisfies the closed
equation~\eqref{eq:eqnQ2}. Similarly, we can calculate exactly mean
first passage times between any two regions where $\vartheta(\xb,\pb)$
is constant by using \eqref{eq:3} with $\sigma(\theta)$ replaced by
$\varsigma(\theta) = \sqrt{\upsilon}
\exp(\tfrac12\beta\mathcal{G}(\theta))$, where $\mathcal{G}(\theta)$
is the free energy associated with $\vartheta(\xb, \pb)$,
\begin{equation}
  \label{eq:freeL}
  \begin{aligned}
    \mathcal{G}(\theta) & = - \beta^{-1} \ln \average{\delta(\vartheta(\xb, \pb) - \theta)} \\
    &\equiv - \beta^{-1} \ln \int_{\RR^{2N}}\!\!\! \varrho_e(\xb, \pb)
    \delta(\vartheta(\xb, \pb) - \theta) d \xb d \pb,
  \end{aligned}
\end{equation}
and $\upsilon$ is given by
\begin{equation}
  \label{eq:nuL}
  \begin{aligned}
    \upsilon = \beta^{-1} \int_{\RR^{2N}} \partial_{\pb}
    \vartheta(\xb, \pb) \cdot \gamma(\xb)
    \partial_{\pb} \vartheta(\xb, \pb) \varrho_e(\xb, \pb) \\
    \times \delta(\vartheta(\xb, \pb) - \theta) d \xb d \pb.
  \end{aligned}
\end{equation} 
Like $\nu$, this factor is constant (independent of
$\theta$). Finally, by using the gauge transformation
$\tilde\vartheta(\bd{x}, \bd{p}) = h(\vartheta(\bd{x}, \bd{p}))$, it
is easy to see that $\tilde\Theta(t) = h(\Theta(t))$
satisfies~\eqref{eq:11} with $G$ replaced by $\mathcal{G}$ and $m$
replaced by
\begin{equation}
  \label{eq:13}
  \mu(\tilde\theta) = \beta^{-1} \average{ \partial_{\pb}
    \tilde\vartheta(\xb, \pb) 
    \cdot \gamma(\xb) 
    \partial_{\pb} \tilde \vartheta(\xb, \pb) \!\mid\! \tilde\vartheta(\xb, \pb) =
    \tilde\theta }.
\end{equation}

Another generalization involves introducing vector-valued collective
variables. In the overdamped case~\footnote{%
  A similar construction holds in the Langevin case and is omitted for
  the sake of brevity}, this can be done by picking $M+1$ points
$\ab_i$, $i = 0, \ldots, M$ and defining $M$ potentials
via~\footnote{%
  Note that $\ab_0$ is used as a reference point here, but it is not
  special in any way: if we relabel the points $\ab_i$, $i = 0,
  \ldots, M$, the new set of potentials can be related to the ones in
  the original labeling by a simple linear transformation.}
\begin{equation}
  \label{eq:potvec}
  \nabla \cdot (\rho_e(\xb) D(\xb) \nabla \theta_i(\xb)) = 
  \tau(\delta(\xb - \ab_i) - \delta(\xb - \ab_0))
\end{equation}
for $i = 1, \ldots, M$. The components of $\Thetab(t) = \thetab(\xb(t)) =
(\theta_1(\xb(t)),$ $\theta_2(\xb(t)),$ $\ldots, \theta_M(\xb(t)))$ then satisfy
the equivalent of~\eqref{eq:eqnQ}
\begin{equation}
  \label{eq:vectval0}
  \dot{\Theta}_i(t) = \sqrt{2} \nabla \theta_i(\xb(t)) \cdot
  D^{1/2}(\xb(t)) \etab(t),
\end{equation}
Like~\eqref{eq:eqnQ}, each of these equations is driftless. However,
unlike~\eqref{eq:eqnQ}, they cannot all be put in a form equivalent
to~\eqref{eq:eqnQ2} by a single rescaling of time: that is because
such a rescaling can only act on one equation in the system at a time,
and cannot be made globally for the all system. We can, however, do
such a rescaling on the equation for any linear combination of the
$\Theta_i(t)$, i.e. on the equation for $ \sum_{i=1}^N c_i \Theta_i(t)
$ where $c_i$, $i=1,\ldots, M$ are arbitrary constants. As a result,
we can calculate exactly the mean first passage time between any two
regions where $\sum_{i=1}^N c_i\theta_i(\xb)$ is constant (again for
an arbitrary set of $c_i$'s) by using the closed system of equations
obtained by averaging~\eqref{eq:vectval0} (compare~\eqref{eq:3}):
\begin{equation}
  \label{eq:vectval}
  \dot{\Theta}_i(t) = \sqrt{2} \sum_{j=1}^M \sigma_{ij}(\Thetab(t))
  \eta_j(t),\quad i=1,\ldots, M
\end{equation}
Here $\eta_j(t)$ for $j=1,\ldots, M$ are independent white-noise
processes, and the entries $\sigma_{ij}(\thetab)$ are defined via 
\begin{equation}
  \label{eq:sigmatensor}
    \sum_{k=1}^M \sigma_{ik}(\thetab)\sigma_{kj}(\thetab)= \nu_{ij} e^{\beta G(\thetab)},
\end{equation}
where $G(\thetab)$ is the free energy associated with $\thetab$:
\begin{equation}
  \begin{aligned}
    G(\thetab) & = - \beta^{-1} \ln \average{\delta(\thetab(\xb) - \thetab)} \\
    & \equiv - \beta^{-1} \ln \int_{\RR^N} \rho_e(\bd{x}) \delta(\thetab(\xb) - \thetab) d \bd{x},
  \end{aligned}
\end{equation}
and $\nu_{ij}$ are the constants given by
\begin{equation}
  \nu_{ij} = \int_{\RR^N} \nabla \theta_i(\xb) \cdot D(\xb) \nabla
  \theta_j(\xb) 
  \rho_e(\bd{x}) \delta(\thetab(\xb) - \thetab) d \bd{x}.
\end{equation}
If we let $\tilde \Thetab = \hb (\Thetab)$, where $\hb: \RR^M\to\RR^M$
is a one-to-one map, \eqref{eq:vectval} becomes (compare~\eqref{eq:11})
\begin{equation}
  \label{eq:14}
  \begin{aligned}
    &\Dot{\tilde \Theta}_i(t) = -\beta\sum_{j=1}^M m_{ij}(\tilde
    \Theta (t))
    \partial_{\theta_j} \tilde G(\tilde\Thetab (t)) \\[-8pt]
    & \quad + \sum_{j=1}^M \partial_{\theta_j} m_{ij}(\tilde \Thetab
    (t)) + \sqrt{2} \sum_{j=1}^M g_{ij}(\tilde \Thetab (t))\,
    \eta_j(t)
  \end{aligned}
\end{equation}
where $\tilde G(\tilde \theta)$ is the free energy associated with
$\tilde \thetab(\xb) = \hb(\thetab(\xb))$, the entries
$m_{ij}(\tilde\thetab)$ are defined as (compare~\eqref{eq:12})
\begin{equation}
  \label{eq:15}
  m_{ij} (\tilde\thetab) =  \bigl\langle \nabla
  \tilde\theta_i(\xb)\cdot 
  D(\xb) \nabla \tilde\theta_j(\xb) \mid
  \tilde\thetab(\xb) = \tilde\thetab  \bigr\rangle.
\end{equation}
and $g_{ij}(\tilde \thetab)$ satisfies $\sum_{k=1}^M
g_{ik}(\tilde\thetab)g_{kj}(\tilde\thetab)= m_{ij}(\tilde\thetab)$.
The result above is not a complete generalization to vector-valued
collective variables since such a generalization should permit to
compute exactly mean first passage times between any regions where
each $\theta_i(\xb)$ takes independent constant values rather than
those where $\sum_{i=1}^N c_i \theta_i(\xb)= cst$. Yet, the ability to
pick the $c_i$'s arbitrarily in this expression (and the $a_i$'s
in~\eqref{eq:potvec}) still offers a lot of flexibility in the range
of regions between which mean first passage times can be calculated
exactly.

Let us end this communication with a few comments about the practical
implications of our results. While it is conceptually pleasing that we
can perform exact dynamical coarse-graining with collective variables
that are not slow (and thereby break free from the limitations of the
standard approach based on Markovian approximation to MZ), the
calculation of these variables involves solving~\eqref{eq:sbke} or
\eqref{eq:sbkeL}, which is by no means straightforward. Techniques such
as transition path
sampling~\cite{bolhuis2002transition,dellago2009transition} or the
string method~\cite{e2002string,e2005finite,maragliano2006string}
could be used for this purpose. Alternatively, our results could be
used to test the quality of putative collective variables. For
example, the method proposed in~\cite{peters2013reaction}, which test
whether a collective variable is Markovian in physical time (which, in
general, requires that it be a slow variable), could be generalized to
test for Markovianity after time rescaling, like in~\eqref{eq:3}
(which requires that the collective variable approximates a potential
$\theta(\xb)$ or $\vartheta(\xb,\pb)$ but not that it be slow). In
some sense, this approach is already at the core of optimal
milestoning~\cite{VaVeCiEl2008,VaVe2008}, but it certainly could be
developed further, and also used to analyze simulation or experimental
data in ways alternative to those proposed
e.g. in~\cite{henry2013comparing,best2013native,kalgin2013new}.

We thank A. Szabo and A. M. Berezhkovskii for interesting
discussions. The research of J. L. was supported in part by the Alfred
P.~Sloan Foundation and the NSF grant DMS-1312659. The research of
E.V.-E. was supported in part by NSF grant DMS07-08140 and ONR grant
N00014-11-1-0345.

\bibliographystyle{spphys}
\bibliography{coarse}

\begin{thebibliography}{10}
\providecommand{\url}[1]{{#1}}
\providecommand{\urlprefix}{URL }
\expandafter\ifx\csname urlstyle\endcsname\relax
  \providecommand{\doi}[1]{DOI \discretionary{}{}{}#1}\else
  \providecommand{\doi}{DOI \discretionary{}{}{}\begingroup
  \urlstyle{rm}\Url}\fi

\bibitem{ayton2007multiscale}
G.S. Ayton, W.G. Noid, G.A. Voth, Curr. Opin. Struct. Biol. \textbf{17}(2), 192
  (2007)

\bibitem{noid2013perspective}
W.~Noid, J. Chem. Phys. \textbf{139}(9), 090901 (2013)

\bibitem{weinan2007heterogeneous}
W.~E, B.~Engquist, X.~Li, W.~Ren, E.~Vanden-Eijnden, Commun. Comp. Phys.
  \textbf{2}(3), 367 (2007)

\bibitem{best2012atomistic}
R.B. Best, Curr. Opin. Struct. Biol. \textbf{22}(1), 52 (2012)

\bibitem{clementi2008coarse}
C.~Clementi, Curr. Opin. Struct. Biol. \textbf{18}(1), 10 (2008)

\bibitem{tozzini2005coarse}
V.~Tozzini, Curr. Opin. Struct. Biol. \textbf{15}(2), 144 (2005)

\bibitem{socci1996diffusive}
N.~Socci, J.N. Onuchic, P.G. Wolynes, J. Chem. Phys. \textbf{104}(15), 5860
  (1996)

\bibitem{best2006diffusive}
R.B. Best, G.~Hummer, Phys.\ Rev.\ Lett. \textbf{96}(22), 228104 (2006)

\bibitem{krivov2006one}
S.V. Krivov, M.~Karplus, J. Phys. Chem. B \textbf{110}(25), 12689 (2006)

\bibitem{bryngelson1995funnels}
J.D. Bryngelson, J.N. Onuchic, N.D. Socci, P.G. Wolynes, Proteins: Structure,
  Function, and Bioinformatics \textbf{21}(3), 167 (1995)

\bibitem{dobson1998protein}
C.M. Dobson, A.~{\v{S}}ali, M.~Karplus, Angewandte Chemie International Edition
  \textbf{37}(7), 868 (1998)

\bibitem{shea2001folding}
J.E. Shea, C.L. Brooks~III, Ann.\ Rev.\ Phys.\ Chem. \textbf{52}(1), 499 (2001)

\bibitem{grabert1982projection}
H.~Grabert, \emph{Projection operator techniques in nonequilibrium statistical
  mechanics}, vol.~95 (Springer-Verlag Berlin, 1982)

\bibitem{chorin2009stochastic}
A.J. Chorin, O.H. Hald, \emph{Stochastic tools in mathematics and science}
  (Springer, 2009)

\bibitem{Zwanzig:1961}
R.~Zwanzig, Phys. Rev. \textbf{124}(4), 983 (1961)

\bibitem{hijon2010mori}
C.~Hij{\'o}n, P.~Espa{\~n}ol, E.~Vanden-Eijnden, R.~Delgado-Buscalioni, Faraday
  Discussions \textbf{144}, 301 (2010)

\bibitem{pavliotis2008multiscale}
G.~Pavliotis, A.~Stuart, \emph{Multiscale methods: averaging and
  homogenization} (Springer, 2008)

\bibitem{EVa2004}
W.~E, E.~Vanden-Eijnden, in \emph{{Multiscale Modelling and Simulation, Ed.
  Attinger, S. and Koumoutsakos, P.}}, ed. by S.~Attinger, P.~Koumoutsakos,
  Lecture Notes in Computational Science and Engineering (Springer, 2004), pp.
  35--68

\bibitem{dellago2009transition}
C.~Dellago, P.G. Bolhuis, in \emph{Advanced computer simulation approaches for
  soft matter sciences III} (Springer, 2009), pp. 167--233

\bibitem{hartmann2013characterization}
C.~Hartmann, R.~Banisch, M.~Sarich, T.~Badowski, C.~Sch{\"u}tte, Entropy
  \textbf{16}(1), 350 (2013)

\bibitem{du1998transition}
R.~Du, V.S. Pande, A.Y. Grosberg, T.~Tanaka, E.S. Shakhnovich, J. Chem. Phys.
  \textbf{108}(1), 334 (1998)

\bibitem{hummer2003transition}
G.~Hummer, J. Chem. Phys. \textbf{120}(2), 516 (2003)

\bibitem{e2005transition}
W.~E, W.~Ren, E.~Vanden-Eijnden, Chem.\ Phys.\ Lett. \textbf{413}, 242 (2005)

\bibitem{EVa2006}
W.~E, E.~Vanden-Eijnden, J. Stat. Phys. \textbf{123}, 503 (2006)

\bibitem{Va2006}
E.~Vanden-Eijnden, in \emph{Computer Simulations in Condensed Matter: from
  Materials to Chemical Biology}, ed. by M.~Ferrario, G.~Ciccotti, K.~Binder,
  Lecture Notes in Physics (Springer, 2006), pp. 453--493

\bibitem{EVa2010}
W.~E, E.~Vanden-Eijnden, Ann.\ Rev.\ Phys.\ Chem. \textbf{61}, 391 (2010)

\bibitem{Elber:04}
A.~Faradjian, R.~Elber, J. Chem. Phys. \textbf{120}, 10880 (2004)

\bibitem{VaVeCiEl2008}
E.~Vanden-Eijnden, M.~Venturoli, G.~Ciccotti, R.~Elber, J. Chem. Phys.
  \textbf{129}, 174102 (2008)

\bibitem{VaVe2008}
E.~Vanden-Eijnden, M.~Venturoli, J. Chem. Phys. \textbf{130}, 194101 (2008)

\bibitem{BerezhkovskiiSzabo:13}
A.M. Berezhkovskii, A.~Szabo, J. Phys. Chem. B \textbf{117}, 13115  (2013)

\bibitem{krivov2012reaction}
S.V. Krivov, J. Chem. Theory Comput. \textbf{9}(1), 135 (2012)

\bibitem{berezhkovskii2004one}
A.~Berezhkovskii, A.~Szabo, J. Chem. Phys. \textbf{122}(1), 014503 (2004)

\bibitem{bass1998diffusions}
R.F. Bass, \emph{Diffusions and elliptic operators} (Springer, 1998)

\bibitem{LeLe2010}
T.~Lelievre, F.~Legoll, Nonlinearity \textbf{23}, 2131 (2010)

\bibitem{maragliano2006string}
L.~Maragliano, A.~Fischer, E.~Vanden-Eijnden, G.~Ciccotti, J.\ Chem.\ Phys.
  \textbf{125}(2), 024106 (2006)

\bibitem{best2010coordinate}
R.B. Best, G.~Hummer, Proc.\ Natl.\ Acad.\ Sci.\ USA \textbf{107}(3), 1088
  (2010)

\bibitem{berezhkovskii2011time}
A.~Berezhkovskii, A.~Szabo, J. Chem. Phys. \textbf{135}(7), 074108 (2011)

\bibitem{bolhuis2002transition}
P.~Bolhuis, D.~Chandler, C.~Dellago, P.~Geissler, Ann.\ Rev.\ Phys.\ Chem.
  \textbf{53}(1), 291 (2002)

\bibitem{e2002string}
W.~E, W.~Ren, E.~Vanden-Eijnden, Phys.\ Rev.\ B. \textbf{66}, 052301 (2002)

\bibitem{e2005finite}
W.~E, W.~Ren, E.~Vanden-Eijnden, J.\ Phys.\ Chem.\ B \textbf{109}, 6688 (2005)

\bibitem{peters2013reaction}
B.~Peters, P.G. Bolhuis, R.G. Mullen, J.E. Shea, J. Chem. Phys.
  \textbf{138}(5), 054106 (2013)

\bibitem{henry2013comparing}
E.R. Henry, R.B. Best, W.A. Eaton, Proc.\ Natl.\ Acad.\ Sci.\ USA
  \textbf{110}(44), 17880 (2013)

\bibitem{best2013native}
R.B. Best, G.~Hummer, W.A. Eaton, Proc.\ Natl.\ Acad.\ Sci.\ USA
  \textbf{110}(44), 17874 (2013)

\bibitem{kalgin2013new}
I.V. Kalgin, A.~Caflisch, S.F. Chekmarev, M.~Karplus, J. Phys. Chem. B
  \textbf{117}(20), 6092 (2013)

\end{thebibliography}

\end{document}